      [1999/12/01 v1.4c Il Nuovo Cimento]
\documentclass[varenna]{cimento}
\usepackage{psfig}  

\title{Projected BCS Wave Functions for Low Dimensional Frustrated Spin Systems}
\author{Luca Capriotti,}
\institute{Kavli Institute for Theoretical Physics and Department of Physics, 
University of California, Santa Barbara, CA 93106}
\author{Federico Becca,}
\institute{Istituto Nazionale per la Fisica della Materia and Universit\'e de Lausanne,
CH-1015 Lausanne, Switzerland}
\author{Alberto Parola,}
\institute{Istituto Nazionale per la Fisica della Materia and  Universit\`a dell'Insubria, 
I-22100 Como, Italy}
\author{Sandro Sorella}
\institute{Istituto Nazionale per la Fisica della Materia, and SISSA, I-34014 Trieste, Italy}          
\begin{document}

\maketitle
%

Non-magnetic ground states are a fascinating possibility allowed by the
physics of quantum antiferromagnets. These states -- which lack
the classical periodical long-range order -- can be stabilized whenever
reduced dimensionality, a small spin value, and/or the presence of competing
interactions lead to strong enough zero-point quantum fluctuations.

One dimensional or quasi-one dimensional spin-half Heisenberg antiferromagnets 
often have non-magnetic ground states.
Indeed for the 1D nearest-neighbors 
Heisenberg model
\begin{equation} \label{heisenbergmodel}
\hat{\cal{H}}=J{_1}\sum_{n.n.}
\hat{{\bf {S}}}_{i} \cdot \hat{{\bf {S}}}_{j}
\end{equation}
the exact solution due to Bethe \cite{bethe} predicts
the absence of {\em true} long-range antiferromagnetic order
even if the ground state is very close to have a 
broken symmetry, with a gapless excitation spectrum 
and a power-law decay of spin-spin correlations.
This is also the case for any array consisting
of an odd number of chains (odd-leg ladder systems).
The ground state of two chains or 
in general of any even-leg ladder system is non-magnetic too.
However -- in contrast to the previous cases -- here
the correlation length is finite and the spectrum is gapped.\cite{twoleg} 
Such a gap is known to decrease exponentially with the number of 
legs \cite{morandi} leading to a gapless 
spectrum in the two dimensional limit, 
where the ground state of the Heisenberg model has
genuine long-range antiferromagnetic order.\cite{manousakis}

Competing interactions may
allow in principle the stabilization of a non-magnetic ground state 
even in truly two dimensional systems.
One of the simplest examples of these {\em frustrated}
systems, which has been also recently realized experimentally \cite{carretta}, 
is the so-called $J_1{-}J_2$ model \cite{chandra,caprioreview} 
\begin{equation} \label{j1j2ham}
\hat{\cal{H}}=J{_1}\sum_{n.n.}
\hat{{\bf {S}}}_{i} \cdot \hat{{\bf {S}}}_{j}
+ J{_2}\sum_{n.n.n.}
\hat{{\bf {S}}}_{i} \cdot \hat{{\bf {S}}}_{j}~,
\end{equation}
where the antiferromagnetic alignment between neighboring spins (due to
$J_1>0$) is hindered by a next-nearest-neighbors antiferromagnetic coupling 
($J_2>0$).

\begin{figure}
\centerline{\psfig{bbllx=120pt,bblly=310pt,bburx=490pt,bbury=490pt,%
figure=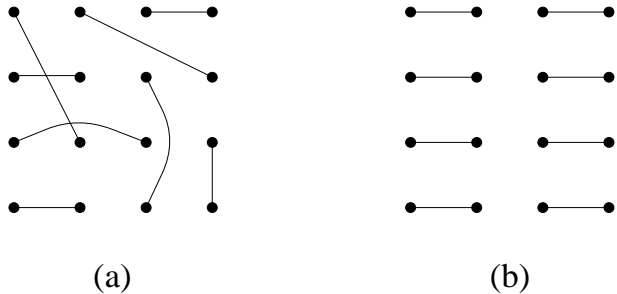,width=90mm,angle=0}}
\caption{\baselineskip .185in
Sketch of a spin liquid (a) and of a symmetry-broken (b) non-magnetic RVB state.
Each stick represents a singlet bond.}\label{fig.rvb}
\end{figure}

Classically, the minimum energy configuration of the 2D  $J_1{-}J_2$ model has the conventional
N\'eel order with magnetic wave vector ${\bf Q}=(\pi,\pi)$ for $J_2/J_1<0.5$. 
Instead for $J_2/J_1>0.5$ the minimum energy configuration is the so-called {\em collinear} 
state with the spins ferromagnetically aligned in one direction and antiferromagnetically
in the other, corresponding to magnetic wave vectors
${\bf Q}=(\pi,0)$ or ${\bf Q}=(0,\pi)$.\cite{chandra2}
Exactly at $J_2/J_1=0.5$ any classical state having zero total spin on each elementary square
plaquette is a minimum of the total energy. These states include both the N\'eel
and the collinear states but also many others with no long-range order so that the occurrence 
of a non-magnetic ground state in the quantum case,  for a small spin value, 
is likely around this value of the $J_2/J_1$ ratio.
Indeed, at present there is a general consensus on
the fact that the combined effect of frustration and zero-point motion
leads to the disappearance of the long-range antiferromagnetic
order marked by the opening of a finite spin gap 
for $ \sim 0.4 < J_2/J_1 < \sim 0.55$.\cite{plaquetto,russi}
The nature of this non-magnetic ground state 
is one of the most interesting puzzles of the physics of frustrated 
spin systems. 
In particular an open question is whether the ground state of the
$J_{1}{-}J_{2}$ Heisenberg model is
a homogeneous spin liquid, i.e., a state with all the symmetries of the Hamiltonian,
as it was originally suggested by Figueirido {\em et al.} \cite{liquid}.
The other possibility is a ground state 
which is still ${\cal SU}(2)$ invariant,
but nonetheless breaks some crystal symmetries, 
dimerizing in some special pattern (see below).
\cite{dagotto,gelfand,read,zith,singh,russi}

\begin{figure}
\centerline{\psfig{bbllx=25pt,bblly=350pt,bburx=560pt,bbury=665pt,%
figure=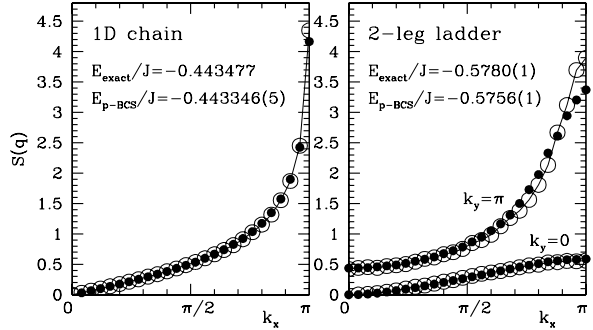,width=80mm,angle=0}}
\caption{\baselineskip .185in \label{sqchleg}
Variational estimate of the magnetic structure factor
for the spin-half Heisenberg chain and two-leg ladder (filled circles).
Empty dots are the numerically exact results obtained with
the Green's function Monte Carlo method.\protect\cite{superb}
}
\end{figure}

A simple picture of a non-magnetic ground state can be given in terms
of the so-called {\em Resonating Valence Bond} (RVB) states.\cite{fazekas}
These are linear superpositions of valence bond states
in which each spin forms a singlet bond
with another spin on the opposite sublattice (say $A$ and $B$) \cite{liang}
\begin{equation}
|\psi_{RVB}\rangle =
\sum_{i_\alpha\in A,j_\beta\in B} h(r_1)\dots h(r_{\frac{N}{2}})
\,\,\, (i_1\,j_1) \dots (i_{\frac{N}{2}},j_{\frac{N}{2}})~,
\label{eq.rvb2}
\end{equation}
where $N$ is the number of sites of the lattice, 
$r_m$ is the distance between the spins forming the $m^{th}$ singlet bond
$(i_m\,j_m)$, and $h(r_m)$ is a bond weight factor.
These states form in general a (overcomplete) basis
of the $S=0$ subspace so that any singlet wave function can be represented in terms of them.
However, they represent a non-magnetic state whenever the short-ranged
bonds dominate the superposition (\ref{eq.rvb2}).
More precisely, it has been numerically shown  by
Liang, Doucot and  Anderson\cite{liang}
that the RVB state (\ref{eq.rvb2}) has no long-range antiferromagnetic order
for bonds that decay as rapidly as $h(r)\sim r^{-p}$,
with $p\ge 5$.
Such bonds can be either homogeneously spatially distributed on the
lattice, with short-range correlations among each other (spin liquid)
[fig.~\ref{fig.rvb} (a)],
or they can break some symmetries of the Hamiltonian, with the dimers frozen
in some special patterns [fig.~\ref{fig.rvb} (b)] as originally predicted for
the $J_1{-}J_2$ model in the regime of strong frustration. 
\cite{dagotto,gelfand,read,zith,singh,russi}

In a seminal paper,\cite{anderson} Anderson proposed 
that a physically transparent description of a RVB state 
can be obtained in fermionic representation
by starting from a BCS-type pairing wave function, of the form
\begin{equation} \label{wavefunction}
|\psi_{\rm BCS}\rangle = {\rm exp} \left( \sum_{i,j}
f_{i,j} \hat{c}^{\dag}_{i,\uparrow} \hat{c}^{\dag}_{j,\downarrow} \right)
|0\rangle~.
\end{equation}                                 
This wave function is the ground state of the well-known BCS Hamiltonian
\begin{equation} \label{hbcs}
H_{BCS}= \sum_{k,\sigma} \epsilon_k c^{\dag}_{k,\sigma} c_{k,\sigma}
           + \sum_k (\Delta_k  c^{\dag}_{k,\uparrow} c^{\dag}_{-k,\downarrow}
           + h.c. )
\end{equation}
where $\sigma=\uparrow,\downarrow$, $\epsilon_k=-2 [\cos k_x + \cos k_y]$ is the free-electron dispersion
and $\Delta_k=\Delta_{-k}$ is the (real) gap function, provided
the Fourier transform $f_k$ of the pairing function, $f_{i,j}$, satisfies: 
$f_k = \Delta_k/(\epsilon_k + \sqrt{\epsilon_k^2 +\Delta_k^2})$.
The non-trivial character of this wave function emerges when we restrict 
to the subspace of fixed number of electrons (equal to the number of sites) and
enforce Gutzwiller projection onto the subspace with no double occupancies: 
singlet pairs do not overlap in real space and this 
wave function can be described by a superposition of 
valence bond states of the form (\ref{eq.rvb2}).\cite{anderson,gros,poilblanc} 

\begin{figure}
\centerline{\psfig{bbllx=65pt,bblly=195pt,bburx=500pt,bbury=710pt,%
figure=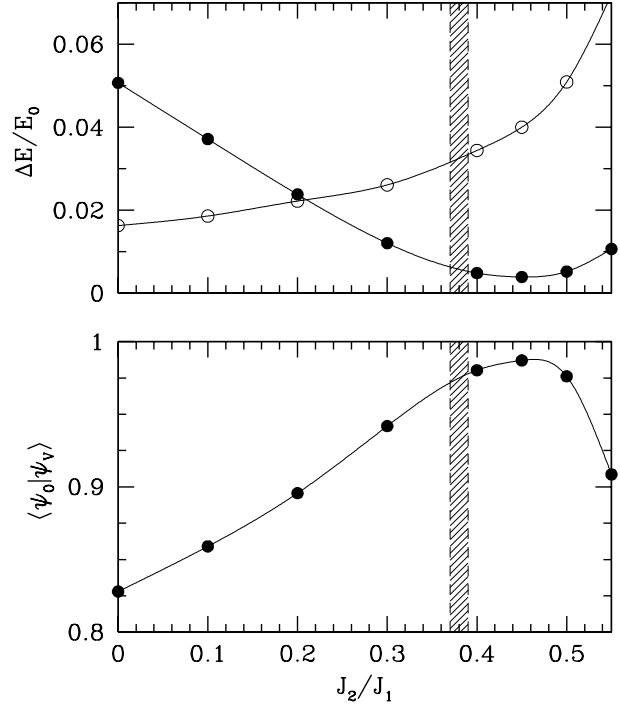,width=80mm,angle=0}}
\caption{\baselineskip .185in \label{overtalk}
Accuracy of the ground-state energy, and overlap between the ground state and the {\em p-}BCS
state (full dots) as a function of $J_2/J_1$, for $N=6\times 6$.
Empty dots are the energy accuracy of
a N\'eel ordered spin-wave wave function.\protect\cite{franjo,caprioreview}
Lines are guides for the eye and the shaded region indicates
the location of the expected transition point  to the non-magnetic phase.
}
\end{figure}

This {\em projected-}BCS ({\em p-}BCS) wave function turns out to be an almost exact representation
of several low-dimensional spin system with non-magnetic ground states.
For instance it provides an excellent variational ansatz of
the ground state of the Heisenberg chain and of the two-leg ladder
giving a very accurate estimate of the ground-state energy (fig.~\ref{sqchleg})
and reproducing almost exactly the antiferromagnetic correlations. 
In the first case the spin structure factor
$S(q)=\langle \hat{{\bf{S}}}_{q} \cdot \hat{{\bf{S}}}_{-q} \rangle$
shows a cusp at $q=\pi$ while for two-leg ladders it has a broad maximum at $q=(\pi,\pi)$.
These features are remarkably
well reproduced by the ({\em p-}BCS) variational wave function (fig.~\ref{sqchleg}), which
generates robust antiferromagnetic correlations at short distances with a 
very simple parameterization
of the gap function: $\Delta_k= \Delta_1 \cos k + \Delta_2 \cos 3k$ for the chain
and $\Delta_k= \Delta_x \cos k_x + \Delta_y \cos k_y$ for the ladder.\cite{superb}

\begin{figure}
\centerline{\psfig{bbllx=50pt,bblly=450pt,bburx=510pt,bbury=680pt,%
figure=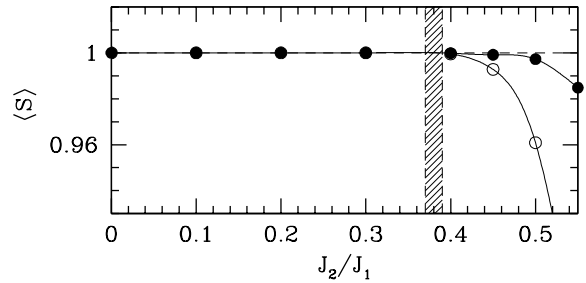,width=80mm,angle=0}}
\caption{\baselineskip .185in \label{signtalk}
Average sign of the {\em p-}BCS state (full dots) as a function of
$J_2/J_1$, for $N=6\times 6$.
Empty dots are the Marshall sign obtained with the {\em p-}BCS state with only
the $d_{x^2-y^2}$ component of the gap function.
Lines are guides for the eye and the shaded region indicates
the location of the expected transition point  to the non-magnetic phase.
}
\end{figure}

In two dimensions this wave function has been already studied 
for the pure Heisenberg model by several authors\cite{gros,poilblanc}
for $\Delta_k \propto  (\cos k_x - \cos k_y)$. In this case it provides a reasonable
prediction for the ground-state energy but it fails in reproducing correctly 
the long-range antiferromagnetic order of the ground state. 
Here we show that this type of RVB state represents an  extremely
accurate variational ansatz for the $J_1{-}J_2$ model in the non-magnetic phase
when the gap function  $\Delta_{k}$ is carefully parameterized. 
In particular,
a definite symmetry is guaranteed to the {\em p-}BCS
state provided $\Delta_{k}$ transforms according to a one dimensional 
representation of the spatial symmetry group. A careful analysis
\cite{rainbow,therevenge}
shows that the odd  component 
of the gap function $\Delta_{k}=-\Delta_{k+(\pi,\pi)}$
may have spatial symmetries different from 
those of the even component  $\Delta_{k}=\Delta_{k+(\pi,\pi)}$. 
Indeed, the best variational energy is obtained when 
the former has $d_{x^2-y^2}$ symmetry, whereas the latter either vanishes or it
has $d_{xy}$ symmetry. 
In order to determine the best variational wave function of
this form we have used a recently developed quantum Monte Carlo
technique \cite{flst} that allows to optimize a large number
of variational parameters with modest computational effort.

The remarkable accuracy of the {\em p-}BCS wave function in describing
the ground state of the 2D $J_1{-}J_2$ model in the regime of 
strong frustration can be shown by calculating the variational 
energy and the overlap with the exact ground state, $|\psi_0\rangle$,
for the largest square cluster $N=6\times6$ where the 
solution can be numerically determined by exact diagonalization.
As shown in fig.~\ref{overtalk} the accuracy of the {\em p-}BCS wave function
rapidly increase by increasing the frustration ratio
$J_2/J_1$ whereas conventional N\'eel ordered spin-wave wave functions
\cite{franjo,caprioreview} quickly become less and less accurate.
Entering the regime of strong frustration $J_2/J_1 \sim 0.45 \pm 0.05$,
where a gapped non-magnetic ground state is expected, the {\em p-}BCS 
wave function becomes impressively accurate with a relative accuracy on the
ground-state energy of order $\sim 4\times 10^{-3}$ and an overlap to the exact 
ground state of $\sim 99\%$,
both improved by more than an order of magnitude with respect 
to the $J_2=0$ case.  This fact implies that the ground state in the strongly
frustrated regime is almost exactly reproduced by a RVB wave function.

\begin{figure}
\centerline{\psfig{bbllx=40pt,bblly=195pt,bburx=510pt,bbury=660pt,%
figure=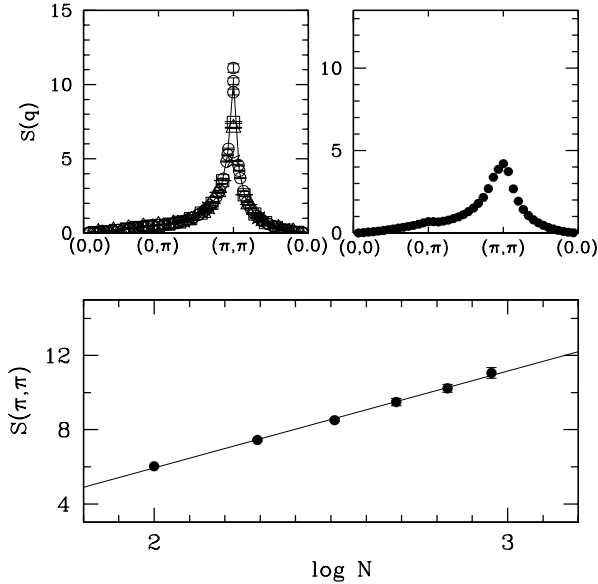,width=80mm,angle=0}}
\caption{\baselineskip .185in \label{sqdxy}
Antiferromagnetic structure factor of the
$d_{x^2-y^2}$ (top left panel) and $d_{x^2-y^2}+ d_{xy}$ (top right panel)
{\em p-}BCS wave functions. Lower panel: size scaling of 
$S(\pi,\pi)$ for the $d_{x^2-y^2}$ {\em p-}BCS state.
}
\end{figure}

Interestingly, the transition to the regime of strong frustration is 
marked by the stabilization at the variational level of a
non-zero $d_{xy}$ component of the gap function.
This allows to reproduce correctly the phases of the 
actual ground-state configurations as
illustrated in fig.~\ref{signtalk}.
A measure of the accuracy of a variational wave function $|\psi_V\rangle$ 
in reproducing the phases of the ground state  can be given in terms of the average sign
$\langle S\rangle =  \sum_x |\langle x|\psi_{\rm V}\rangle|^2
{\rm Sgn} \big[ \langle x|\psi_{\rm V}\rangle \langle x|\psi_0\rangle \big]~.
$
In the unfrustrated case it is well known that such phases are
determined by the so-called Marshall sign rule \cite{marshall}:
on each real space configuration $|x\rangle$,
the sign of the ground-state wave function
is determined only by the number of spin down in one of the two sublattices.    
This feature, rigorously valid for $J_2=0$, 
turns out to be a very robust property
for weak frustration ($J_2/J_1 < \sim0.3$).\cite{richter} 
However, it is clearly violated when frustration plays an important role.
It can be shown \cite{rainbow,therevenge} that the Marshall sign (i.e., $\langle S \rangle =1$ for $J_2/J_1=0$)
can be obtained using the {\em p-}BCS wave function, with only the 
$d_{x^2-y^2}$ component, so that this wave function 
provides an almost exact representation of the ground-state phases for weak frustration.
However, for $J_2/J_1 > \sim 0.4$, the phases of the
wave function are considerably affected by the strong frustration
and only when a sizable $d_{xy}$ component is stabilized at 
the variational level, this property
can be correctly reproduced.

\begin{figure}
\centerline{\psfig{bbllx=40pt,bblly=260pt,bburx=500pt,bbury=645pt,%
figure=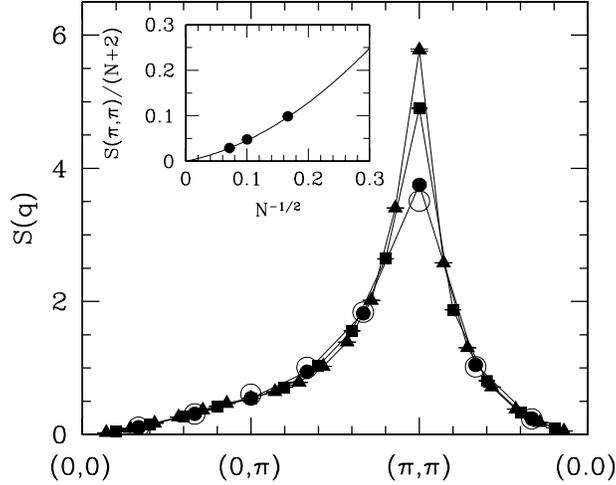,width=80mm,angle=0}}
\caption{\baselineskip .185in \label{sq}
Variational estimate of $S(q)$ for (from the lower to the upper curve) $N=6\times 6$, $10\times 10$, and $14
\times 14$.
Large empty circles are the exact diagonalization results for the $N=6\times6$.
Inset: size-scaling of the order parameter squared.
}
\end{figure}

An even more remarkable effect associated with the $d_{xy}$ component of the
gap function is the change induced on antiferromagnetic correlations. 
As it is shown, in fig.~\ref{sqdxy} the finite-size magnetic structure factor
of the $d_{x^2-y^2}$ {\em p-}BCS state 
is sharply peaked around the antiferromagnetic wave vector ${\bf Q}=(\pi,\pi)$
giving rise to a logarithmic divergence in the thermodynamic limit.
Such a divergence is instead washed out in presence 
of the $d_{xy}$ component of the gap function, leading to a state 
with weaker short-range antiferromagnetic correlations which is of course
more suitable to describe the spin-gapped strongly-frustrated phase.

Of course, the accuracy in the energy does not necessarily 
guarantee a corresponding accuracy in correlation functions. 
However, as shown in fig.~\ref{sq}, the comparison 
of the  magnetic structure factor
with the exact result gives
a clear indication that correlation functions obtained by the variational 
approach are essentially exact.
Furthermore using the stochastically implemented Lanczos technique
and the variance extrapolation method \cite{flst} we have verified 
that the accuracy of the {\em p-}BCS wave function is
preserved by increasing the lattice size.\cite{rainbow}

\begin{figure}
\centerline{\psfig{bbllx=30pt,bblly=130pt,bburx=570pt,bbury=695pt,%
figure=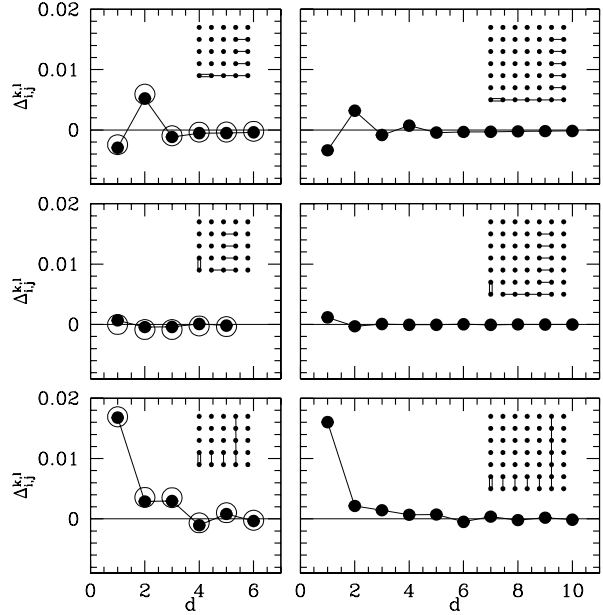,width=80mm,angle=0}}
\caption{\baselineskip .185in \label{dimer}
Variational estimate of the dimer-dimer correlation functions $\Delta_{i,j}^{k,l}$ obtained by keeping
fixed the position of the bond $(i,j)$ (double stick) and
moving the bond $(k,l)$ (single stick) along the indicated patterns.
$d$ is the Manhattan distance. $6\times 6$ (left), $10\times 10$ (right).
Large empty circles are the exact diagonalization results for the $N=6\times6$.
}
\end{figure}

In order to investigate the existence of long-range dimer-like correlations,
as in the columnar 
or the plaquette valence bond state, we have calculated
the dimer-dimer correlation functions, 
$\Delta_{i,j}^{k,l}=\langle \hat{S}_i^z\hat{S}_j^z \hat{S}_k^z \hat{S}_l^z\rangle - \langle \hat{S}_i^z\hat{S}_j^z \rangle \langle \hat{S}_k^z \hat{S}_l^z\rangle$. 
In presence of some broken spatial symmetry, the latter  should converge 
to a finite value for large distance. 
This is clearly ruled out by our results, shown 
in fig.~\ref{dimer}, with a very robust confirmation of the 
{\em liquid}  character of the ground state for $J_2/J_1\simeq 0.5$, which is correctly 
described by our variational approach.

A totally symmetric spin-liquid solution proposed for this 
model in ref.~\cite{liquid} was actually rather unexpected after the work of Read 
and Sachdev,\cite{read} providing arguments in favor of spontaneous 
dimerization. This conclusion was supported by series expansion \cite{singh,russi}
and quantum Monte Carlo studies included the one done  by two of us.\cite{plaquetto} 
It is clear however that it is very hard 
to reproduce a fully symmetric spin liquid ground state, with  any technique, numerical 
or analytical, based on reference states explicitly breaking 
some lattice symmetry.

In conclusion, the spin-liquid RVB ground state, originally proposed to 
explain high-Temperature superconductivity, is indeed a very robust property 
of strongly frustrated low-dimensional spin systems.  
Due to the success in reproducing the non-magnetic ground states 
of other low-dimensional spin systems like the 1D chain and the two-leg ladder, 
\cite{superb}
we expect that the {\em p-}BCS RVB wave function 
represents the  {\em generic}  variational state
for a spin-half spin liquid, once the pairing function $f_{i,j}$ 
is exhaustively parameterized
according to the symmetries of the Hamiltonian.
Work is in progress on this line of research.\cite{therevenge,backtobethe}

\acknowledgments

We thank C. Lhuillier, F. Mila, and D.J. Scalapino for stimulating discussions.
This work has been partially supported by 
MURST (COFIN01). L.C. was supported by NSF grant DMR-9817242.


\end{document}